\documentclass [12pt] {article}
\begin{document}

\title{The standard "static" spherically symmetric ansatz with perfect fluid source revisited}

\author{\.{I}brahim Semiz\thanks{mail: ibrahim.semiz@boun.edu.tr} \\    
   Department of Physics  \\
   Bo\u{g}azi\c{c}i University\\
Bebek, \.{I}stanbul, TURKEY\\}

\date{ }

\maketitle

\begin{abstract}
Considering the standard ''static" spherically symmetric ansatz 
$ds^{2} = -B(r) dt^{2} + A(r) dr^{2} + r^{2} d\Omega^{2}$ 
for Einstein's Equations with perfect fluid source, we ask how we can interpret solutions where $A(r)$ {\em and} $B(r)$ are not positive, as they must be for the static matter source interpretation to be valid.

Noting that the requirement of Lorentzian signature implies \mbox{$A(r) B(r) >0$}, we find two possible interpretations:

(i) The nonzero component of the source four-velocity does {\em not} have to be $u^{0}$. This provides a connection from the above ansatz to the Kantowski-Sachs (KS) spacetimes.

(ii) Regions with negative $A(r)$ and $B(r)$ of  "static" solutions in the literature must be interpreted as corresponding to tachyonic source.

The combinations of source type and four-velocity direction result in four possible cases. One is the standard case, one is identical to the KS case, and two are tachyonic. The dynamic tachyonic case was anticipated in the literature, but the static tachyonic case seems to be new. We derive Oppenheimer-Volkoff-like  equations for each case, and find some simple solutions. We conclude that new "simple" black hole solutions of the above form, supported by a perfect fluid, do not exist.
\end{abstract}

\section{Introduction: Constraints on the signs of metric functions}

Static spherically symmetric perfect fluid (SSSPF) solutions of Einstein's Equations abound in the literature (e.g.~\cite{delgaty&lake}, \cite[Sect.16.1]{exsols}, and their references) because of the relative simplicity of the setting.

But what {\em is} an exact solution? The Einstein's Equations read
\begin{equation}
G_{\mu\nu} = \kappa T_{\mu\nu}
\end{equation}
 and for any metric $\tilde{g}_{\mu\nu}$, an Einstein tensor $\tilde{G}_{\mu\nu}$ can be calculated; so, the metric $\tilde{g}_{\mu\nu}$ can be claimed to solve Einstein's Equations for the source given by the stress-energy-momentum tensor $\tilde{T}_{\mu\nu} = \tilde{G}_{\mu\nu} / \kappa$ (If one wants to include a cosmological constant $\Lambda$ in Einstein's equations, the argument can be modified in an obvious way). Therefore the question of the validity of a solution leads to the question of the acceptability of the required stress-energy-momentum tensor, i.e. existence or nonexistence of matter or fields corresponding to that $T_{\mu\nu}$. General Relativity is of no help here, the intrinsic properties of matter or fields are outside its realm.

Some general requirements for $T_{\mu\nu}$ are proposed, collectively known as {\em energy conditions}  (e.g. \cite{wald}). For example, the weak energy condition states that the energy density should be nonnegative according to every observer. Alternatively, one might choose to impose {\em strong} or {\em dominant} energy conditions. If the source is a perfect fluid, the stress-energy-momentum tensor takes the form
\begin{equation}
T_{\mu\nu} = (\rho + p) u_{\mu} u_{\nu} + p g_{\mu\nu}  \label{pfemt}
\end{equation}
where $\rho$ and $p$ are the energy density and pressure, respectively, as measured by an observer moving with the fluid, and $u_{\mu}$ is its four-velocity. In this case the weak energy condition takes the form $\rho \geq 0$, $\rho+p \geq 0$; the strong energy condition the form $\rho+p \geq 0$, $\rho+3p \geq 0$; the dominant energy condition the form $\rho \geq |p|$. Various tests of acceptability along these lines, such as positivity of energy density and pressure, regularity at origin, subluminal sound speed, etc. are applied to 127 listed candidates for SSSPF solutions in \cite{delgaty&lake}.

But on one hand, singularities are unavoidable in various contexts, notably gravitational collapse \cite{singth}. The singularity theorems prove this by using various energy conditions as prerequisites. On the other hand, negative energy densities are possible in Quantum Field Theory \cite{nonpositivity}.  Examples include the well-known Casimir Effect~\cite{casimir}, squeezed states of light~\cite{sqzd-review, sqzd-negenergy, sqzd-worm} and radiation from moving mirrors~\cite{mirrors}. While uncertainty-principle-like restrictions are suggested on negative energies~\cite{ford}, these phenomena represent breakdown of all energy conditions.  In cosmology, negative pressures are considered routinely since the advent of the concept of inflation~\cite{inflation-star, inflation-guth, inflation-review} in the eighties, and especially Dark Energy \cite{de} in the last decade, after the discovery of the acceleration of the expansion of the universe \cite{acceleration-hiZsst,acceleration-SCP}. Negative energy densities are occasionally considered elsewhere in General Relativity, as well  (e.g. \cite{negenergy}). Therefore, while conditions should be imposed  on $T_{\mu\nu}$ to decide acceptability of metrics as solutions of Einstein's Equations,  it is not very clear what those conditions should be.

Yet, conditions on $T_{\mu\nu}$ are not sufficient; more primary is the correctness of the signature of the metric: Physics should locally be Minkowskian. The Einstein Equations not only do not guarantee correct signature, they even allow signature change -- even some of the earliest static spherically symmetric solutions, the Einstein static universe\footnote{For the form of the solution and reference to the original publication, see \cite{delgaty&lake}}, Tolman IV and Tolman V  \cite{Tolman}, for example, involve expressions that allow different signatures in different ranges of coordinates. Of course, in the Einstein static universe, the maximum value of the coordinate for correct signature is interpreted as the size of the universe, therefore the region in which the signature is wrong can be argued to be irrelevant.   

The standard static spherically symmetric ansatz is
\begin{equation}
ds^{2} = -B(r) dt^{2} + A(r) dr^{2} + r^{2} d\Omega^{2}   \label{ansatz}
\end{equation}
where $d\Omega^{2} = d\theta^{2} + \sin^{2}\theta \, d\phi^{2}
$ is the metric of a two-sphere. In terms of the metric functions of this ansatz, the above requirement of correct signature becomes
\begin{equation}
A(r) B(r) >0,   \label{signature}
\end{equation}
a constraint not always explicitly stated in published solutions.

The requirement that the four-velocity of the source fluid should be real further constrains the signs of the metric functions for SSSPF solutions: When ansatz (\ref{ansatz}) is assumed, it is customary (\cite{delgaty&lake}, \cite[Sect.16.1]{exsols}, or any General Relativity textbook) to take the fluid to be at rest, since the spacetime is static, i.e.
\begin{equation}
u^{\mu} = u^{0} \delta_{0}^{\mu}	\label{staticu}
\end{equation}
The normalization of the four-velocity gives  
\begin{equation}
u_{\mu}u^{\mu} = -1 	\label{unorm}
\end{equation}
Combining,
\begin{equation}
 -B(r) (u^{0})^{2}=-1	\label{b-u02}
\end{equation}
therefore, if $B(r)$ is negative, $u^{0}$ becomes imaginary. This is obviously unacceptable, more so than any properties of the energy density or pressure of the fluid.  {\em Regions in which $B(r)$ is negative should be excluded from the set of SSSPF solutions} (One can also arrive at this conclusion by demanding that the relevant Killing vector be timelike). This constraint is not always explicitly stated either.

In recognition of this requirement, $A(r)$ and $B(r)$ are sometimes written as $e^{\Phi(r)}$ and $e^{\Psi(r)}$, respectively (e.g. \cite{mtw}), but when one carries out the integrals necessary for solution of Einstein's Equations, one often gets logarithms, canceling the exponential (neglecting the requirement that logarithm cannot have a negative argument), and loses the positivity property of $e^{\Phi(r)}$ and $e^{\Psi(r)}$. The best-known example of this is the Schwarzschild (exterior) solution, where  $A(r)$ and $B(r)$ are negative inside the horizon.

Of course, it is this region that makes the Schwarzschild solution a black hole spacetime. But the Schwarzschild solution is a vacuum solution, there is no source fluid, therefore no four-velocity (that would become imaginary). The same is true for the K\"ottler\footnotemark[1] (aka Schwarzschild-de Sitter) solution, if it is taken as a vacuum solution of Einstein's Equations with cosmological constant $\Lambda$; but if it is taken as a solution of the original Einstein Equations with a "static" perfect fluid of equation of state $p=-\rho$  as a source, it {\em does} have the problem of imaginary four-velocity, except in the region between the horizons. The identification and reinterpretation or cure of unacceptable parameter and/or coordinate ranges for SSSPF solutions in the literature will be reported separately \cite{improper}.

How can one interpret regions or solutions with negative $A(r)$ and $B(r)$, if not as SSSPF solutions? We turn to this question in the next section.

\section{Alternatives: Dynamic and/or tachyonic cases} \label{s:alt}

The requirement of positivity of $A(r)$ and $B(r)$ follows from  (\ref{staticu}), (\ref{unorm}) and Lorentzian signature of the metric. Therefore if we try to find meaning for regions of negative $A(r)$ and $B(r)$, we must consider violations of (\ref{staticu}) or (\ref{unorm}). Of course, it is well-known that regions with negative $A(r)$ and $B(r)$ are not really static (hence, the quotes in the title and abstract); calling a coordinate $t$ does not make it timelike, the negative sign of the corresponding metric element does.

We would like to first  point out that (\ref{staticu}) can be violated; that is, the nonzero component of the source four-velocity does {\em not} have to be $u^{0}$. The standard assumption  (\ref{staticu}) that $u^
{0}$ is nonzero is usually imposed {\em ad hoc, before} applying Einstein's Equations;  it is not the only one compatible with them. To see this, consider the nonzero components of the Einstein tensor corresponding to the standard ansatz (\ref{ansatz}) for the metric:
\begin{eqnarray}
G_{00} & = &  \frac{B}{r^{2}} \left( 1- \frac{1}{A} + \frac{rA'}{A^{2}} \right) \\ 
G_{11} & = & \frac{1}{r^{2}} \left(1- A + \frac{rB'}{B}\right)\\ 
G_{22} & = & \frac{r}{2A} \left[ -\frac{A'}{A} + \frac{B'}{B} - \frac{rA'B'}{2 AB} - \frac{rB'^{2}}{2 B^{2}} + \frac{rB''}{B} \right]\\ 
G_{33} & = & G_{22} \sin^{2}\theta  \label{G2233}
\end{eqnarray}
where $A(r)$ and $B(r)$ are written as $A$ and $B$ for brevity, and prime denotes $r$-derivative. The last relation comes from spherical symmetry. Since both $g_{\mu\nu}$ and $G_{\mu\nu}$ are diagonal, Einstein Equations with (\ref{pfemt}) give
\begin{equation}
u_{\mu} u_{\nu} = 0 \;\;\;\;\;\;\;  {\rm for} \;\; \mu \neq \nu     \label{offdiagu}
\end{equation}
if $\rho + p$ is nonzero. We do not consider the special case  $\rho + p=0$ since this leads to the well-known K\"ottler\footnotemark[1] (SdS) solution. Otherwise, eq.~(\ref{offdiagu}) means that only one component of $u_{\mu}$ can be nonzero. Eq. (\ref{G2233}) means $T_{33} = T_{22} \sin^{2}\theta$, giving $(u_{3})^{2} = (u_{2})^{2}  \sin^{2}\theta$, therefore both $u_{2}$ and $u_{3}$ must be zero; but there is no reason why $u_{1}$ should be.

Alternatively, if (\ref{unorm}) is violated, we must replace it with
\begin{equation}
u_{\mu}u^{\mu} = +1    \label{utach}
\end{equation}
since one can always normalize the four-velocity to $\pm 1$. Then the fluid is {\em tachyonic}.

While tachyons would seem to violate causality, tachyonic fields have been put forward as one of the candidates for dark energy (see e.g. \cite{tachDE}).  Also, in \cite{davies} tachyonic particles are considered to provide a certain equation of state. But there, the fluid is still at rest, therefore the fluid four-velocity still obeys (\ref{staticu}) and (\ref{unorm}), it is the random motion of the fluid particles that is tachyonic. Here though, we are forced to consider the fluid itself having tachyonic property. 

So we have a $2 \times 2$ matrix of possibilities: $u_{0}$ or $u_{1}$ may be nonzero, the fluid may be normal or tachyonic. We label the cases with two letter-abbreviations, the first showing the type of source (normal or tachyonic -- N or T), the second showing if the spacetime is static or dynamic (S or D). Incidentally, our ansatz cannot accomodate a null fluid, because 
$-B(r) (u^{0})^{2}$ [in the $u_{0}$ nonzero case] or $A(r) (u^{1})^{2}$  [in the $u_{1}$ nonzero case] cannot be zero without decreasing the dimensionality of the spacetime.

Whatever the case, the Einstein Equations provide three equations for the four unknowns $A(r)$, $B(r)$, $\rho(r)$ and $p(r)$, so some extra input is needed to determine a solution. This extra input can be in the form of a mathematically motivated ansatz for one of $A(r)$ and $B(r)$, chosen so that the other
 one can be easily found\footnote{Actually, one can specify one of $A(r)$ and $B(r)$ {\em arbitrarily}, and get a differential equation for the other. In this sense, all possible solutions are expressible in terms of an arbitrary function, but the resulting nonlinear differential equation is not always solvable analytically.}. But this will lead in general to complicated expressions for $\rho(r)$ and $p(r)$, which may be difficult to interpret physically. A second, physically motivated approach is to implement the properties of the desired source fluid via a relation between its pressure and density, an {\em equation of state} $f(p,\rho)=0$. But this will lead to a nonlinear differential equation, in general hard or impossible to solve analytically.
        
We now turn to the consideration of the four possible cases. 

\subsection{Case NS. $u_{0}$  nonzero, fluid normal} 

This is the well-known case, included here for sake of comparison with the other cases. In this case we have 
\begin{equation}
T_{00} = \rho(r) B(r), \;\;\; T_{11} = p(r) A(r), \;\;\; T_{22} = p(r) r^{2}
\end{equation}
so that Einstein's Equations turn into
\begin{eqnarray}
\frac{1}{r^{2}} \left( 1- \frac{1}{A} + \frac{rA'}{A^{2}} \right)  & = & \kappa \rho    \label{EEns1} \\ 
\frac{1}{r^{2}} \left(1- A + \frac{rB'}{B}\right)   & = & \kappa p A    \label{EEns2}\\ 
\frac{1}{2} \left[ -\frac{A'}{A} - \frac{B'}{B} - \frac{rA'B'}{2 AB} - \frac{rB'^{2}}{2 B^{2}} + \frac{rB''}{B} \right]    & = & \frac{1}{r} \left(1- A \right)      \label{EEns3}
\end{eqnarray}

The mathematically motivated approach \cite{Tolman} consists of putting an ansatz for $A(r)$ or $B(r)$ into (\ref{EEns3}), solving for the other, then finding $\rho(r)$ and $p(r)$ via  (\ref{EEns1}) and (\ref{EEns2}).  It is actually possible to change variables so that (\ref{EEns3}) becomes linear in both dependent variables \cite[Sect. 16.1]{exsols}, \cite{buchdahl,visseretal}.

The physically motivated approach \cite{ov} uses the integrability of the paranthesis in (\ref{EEns1}) to define a function $F(r)$
\begin{equation}
F(r) = \kappa \int \rho r^{2} dr          \label{FDefNS}
\end{equation}
which here is $\kappa/4\pi$ times the "mass function" defined in the literature. Then $B'/B$ is also
expressed in  terms of $F$ via (\ref{EEns2}), and finally substitution for $A$, $B$ and their derivatives in (\ref{EEns3}), gives
\begin{eqnarray}
A & = & \frac{r}{r-F}     \label{EEns4} \\ 
\frac{B'}{B}   & =  & \frac{\kappa p r^{2} + 1}{r-F}  -   \frac{1}{r}   \label{EEns5}\\ 
p'   & = & - \frac{(\kappa p r^{3} + F)}{2 r (r-F)} (\rho + p)     \label{EEns6}
\end{eqnarray}

Eq. (\ref{EEns6}) is the well-known Oppenheimer-Volkoff (OV) equation. In this equation now one would put $p$ in terms of $\rho$ via an equation of state, then $\rho$ in terms of $F'$, via (\ref{FDefNS}), eventually getting a differential equation for $F$. After solving for $F$, $A$ and $B$ would be found via (\ref{EEns4}) and (\ref{EEns5}), giving a metric for that equation of state. {\em One has to recall that solutions are only valid for positive $A(r)$ and $B(r)$.}

\subsection{Case TD. $u_{0}$  nonzero, fluid tachyonic} \label{secTD}

The possibility of tachyonic fluids was pointed out before as a feature of some solutions with timelike radial coordinate \cite{das}, but here we argue that it is generic in two of the four cases compatible with the standard ansatz (\ref{ansatz}).

Because $u_{0}^{2} = -B$ now, in this case we have 
\begin{equation}
T_{00} = -[\rho(r)+2p(r)] B(r), \;\;\; T_{11} = p(r) A(r), \;\;\; T_{22} = p(r) r^{2}
\end{equation}
i.e. $p$ also enters  into the expression for $T_{00}$. Actually, it is not clear what meaning $\rho$ and $p$ would have for a tachyonic fluid, since the energy density $\rho$ and pressure $p$ for a perfect fluid are defined as those that would be measured by an observer moving with the fluid. Nevertheless, we may consider them to be the appropriate functions in a stress-energy-momentum tensor of the form (\ref{pfemt}).

The Einstein's Equations in this case become
\begin{eqnarray}
\frac{1}{r^{2}} \left( 1- \frac{1}{A} + \frac{rA'}{A^{2}} \right)  & = & - \kappa (\rho+2p)    \label{EEtd1} \\ 
\frac{1}{r^{2}} \left(1- A + \frac{rB'}{B}\right)   & = & \kappa p A    \label{EEtd2}\\ 
\frac{1}{2} \left[ -\frac{A'}{A} - \frac{B'}{B} - \frac{rA'B'}{2 AB} - \frac{rB'^{2}}{2 B^{2}} + \frac{rB''}{B} \right]    & = & \frac{1}{r} \left(1- A \right)      \label{EEtd3}
\end{eqnarray}
i.e. the second and third equations are the same as in case NS. If one follows the mathematically motivated approach, formally the same functions for $A(r)$ and $B(r)$ can be used as in case NS, giving even the same $p(r)$, but $\rho(r)$ will be different.  

In the physically motivated approach we must use a different definition for $F(r)$:
\begin{equation}
F_{TD}(r) = - \kappa \int (\rho+2p) r^{2} dr          \label{FDefTD}
\end{equation}
and the equations (\ref{EEns4})-(\ref{EEns6}) are replaced by
\begin{eqnarray}
A & = & \frac{r}{r-F_{TD}}     \label{EEtd4} \\ 
\frac{B'}{B}   & =  & \frac{\kappa p r^{2} + 1}{r-F_{TD}}  -   \frac{1}{r}   \label{EEtd5}\\ 
p'   & = &  \frac{(\kappa p r^{3} + F_{TD})}{2 r (r-F_{TD})} (\rho + p)     \label{EEtd6}
\end{eqnarray}

One might call eq. (\ref{EEtd6}) the tachyonic Oppenheimer-Volkoff equation, and treat it similarly. It looks similar to the corresponding eq. (\ref{EEns6}) of case NS, apart from a sign; but this is misleading: The substitution of $\rho$ in terms of $F'$ via (\ref{FDefTD}) is different, possibly leading to a quite different differential equation for a given equation of state. Alternatively, it can be brought into the same form by the substitution $\tilde{\rho}=-(\rho + 2 p)$, but then the equation of state must be changed to $f(p,-\tilde{\rho}-2p)=0$. 

{\em Any solutions one gets for this case will only be valid for negative $A(r)$ and $B(r)$,} therefore describe dynamic regions. 

\subsection{Case ND (KS). $u_{1}$  nonzero, fluid normal} 

In this case, $u_{1}^{2} = -A$, {\em which also makes $B$ negative} by the signature requirement, and therefore this case also describes dynamic regions. By analogy with the Schwarzschild vacuum solution, we may expect this case to be relevant in gravitational collapse situations in the region {\em after} horizon formation.  

For this case we have 
\begin{equation}
T_{00} = -p(r) B(r), \;\;\; T_{11} = -\rho(r) A(r), \;\;\; T_{22} = p(r) r^{2}
\end{equation}

Then, Einstein's Equations are
\begin{eqnarray}
\frac{1}{r^{2}} \left( 1- \frac{1}{A} + \frac{rA'}{A^{2}} \right)  & = & - \kappa p    \label{EEnd1} \\ 
\frac{1}{r^{2}} \left(1- A + \frac{rB'}{B}\right)   & = & - \kappa \rho A    \label{EEnd2}\\ 
\frac{1}{2} \left[ \frac{A'}{A} + \frac{B'}{B} - \frac{rA'B'}{2 AB} - \frac{rB'^{2}}{2 B^{2}} + \frac{rB''}{B} \right]    & = & \frac{1}{r} \left(1- A \right)      \label{EEnd3}
\end{eqnarray}

Note that although the only nonzero component  of the fluid's four-velocity is $u_{1}$, the fluid is still moving along the timelike coordinate. So, the coordinates are comoving if we relabel $r$ and $t$, and our case ND corresponds to the subcase $Y \rightarrow t$, $e^{2\lambda} \rightarrow -B(t)$, $e^{2\nu} \rightarrow -A(t)$ of the general non-static spherically symmetric perfect fluid equations, in the notation of~\cite[Sect.16.2]{exsols}. Therefore one can see that the rotation and acceleration, two of the quantities used to classify perfect fluid solutions~\cite[sect.15.6.1]{exsols}, vanish by construction in this case.

Shear, however, does not vanish: If we look for a shearfree solution here, we are led to $B(r)=Cr^{2}$. Then (\ref{EEnd3}) gives $A=0$, which is not acceptable. Therefore, all solutions given in Sect. \ref{ndsols} have shear. Expansion in general is nonzero as well, but a nonexpanding solution can also be found (Sect. \ref{ndsols}, Sol. ND5). 

The general equations are difficult to solve for non-vanishing shear~\cite[Sect.16.2]{exsols}, but one might hope this subcase to be easier, since it is covered by the well-treated ansatz (\ref{ansatz}). In fact, this case corresponds to the Kantowski-Sachs (KS) cosmological models (\cite[Sect.15.6.5]{exsols}, \cite{ks} -- since the range of the spacelike coordinate $t$ is infinite). 

We can derive an OV-like formalism for this case as well. Eq. (\ref{EEnd1}) leads to the definition
\begin{equation}
F_{ND}(r) = - \kappa \int p r^{2} dr          \label{FDefND}
\end{equation}
and we get
\begin{eqnarray}
A & = & \frac{r}{r-F_{ND}}     \label{EEnd4} \\ 
\frac{B'}{B}   & =  & \frac{1-\kappa \rho r^{2}}{r-F_{ND}}  -   \frac{1}{r}   \label{EEnd5}\\ 
\rho'   & = &  \frac{3F_{ND}-4r+ \kappa \rho r^{3}}{2 r (r-F_{ND})} (\rho + p).     \label{EEnd6}
\end{eqnarray}

Eq. (\ref{EEnd6}) is mathematically less similar to the OV equation than (\ref{EEtd6}), and physically it is totally different: Instead of change of pressure with depth, it gives change of density with time, since $r$ in this case is timelike.

\subsection{Case TS. $u_{1}$  nonzero, fluid tachyonic} 

Once the possibility of a tachyonic fluid is considered, the case of a static spacetime supported by a such a fluid becomes inevitable; however, this case seems not to have been considered in the literature so far. Of course, the same caveats about the meanings of $\rho$ and $p$ apply as in case TD.  

Now, $u_{1}^{2} = A$, therefore $B$ is also positive. We have
\begin{equation}
T_{00} = -p(r) B(r), \;\;\; T_{11} = (\rho(r)+2p(r)) A(r), \;\;\; T_{22} = p(r) r^{2}
\end{equation}

The Einstein's Equations in this case are
\begin{eqnarray}
\frac{1}{r^{2}} \left( 1- \frac{1}{A} + \frac{rA'}{A^{2}} \right)  & = & - \kappa p    \label{EEts1} \\ 
\frac{1}{r^{2}} \left(1- A + \frac{rB'}{B}\right)   & = &  \kappa (\rho+2p) A    \label{EEts2}\\ 
\frac{1}{2} \left[ \frac{A'}{A} + \frac{B'}{B} - \frac{rA'B'}{2 AB} - \frac{rB'^{2}}{2 B^{2}} + \frac{rB''}{B} \right]    & = & \frac{1}{r} \left(1- A \right)      \label{EEts3}
\end{eqnarray}
i.e. the first and third equations are the same as in case ND. Because of this, the definition of $F$ is also the same:
\begin{equation}
F_{TS}(r) = - \kappa \int p r^{2} dr          \label{FDefTS}
\end{equation}
leading to
\begin{eqnarray}
A & = & \frac{r}{r-F_{TS}}     \label{EEts4} \\ 
\frac{B'}{B}   & =  & \frac{1+\kappa (\rho+2p) r^{2}}{r-F_{TS}}  -   \frac{1}{r}   \label{EEts5}\\ 
\rho' + 2p'  & = &  \frac{3F_{TS}-4r- \kappa (\rho+2p) r^{3}}{2 r (r-F_{TS})} (\rho+p)     \label{EEts6}
\end{eqnarray}
This equation would be best handled by the substitutions $\tilde{\rho}=-(\rho + 2 p)$ and $f(p,-\tilde{\rho}-2p)=0$ mentioned at the end of section~\ref{secTD}.

\section{Some simple solutions} 

In this section, we present some simple solutions for the four cases described above. As mentioned in the beginning of the previous section, an extra ansatz is needed to solve the three equations. The simple mathematical ans\"{a}tze we take are constancy\footnote{The first four terms of the third equation in each case can be written as $
\left(\frac{A'}{A} + \frac{B'}{B}\right) \left(1 \pm \frac{r}{2}\frac{B'}{B}\right)
$, so assuming either $AB=$~Const or $\frac{B'}{B}=\mp \frac{2}{r}$ would seem to be other possible simplifying ans\"{a}tze. However, the first leads to the K\"{o}ttler (SdS) solution\footnotemark[1] again, while the second leads to $A=$~Const, an ansatz we already use.
} of $A(r)$ or $B(r)$; the simple physical ans\"{a}tze are very simple equations of state, such as $p=0$ (dust in cases NS and ND), $\rho=$~Const ("incompressible" fluid), maybe $p=$~Const.  We give the functions $A(r)$, $B(r)$, $p(r)$ and $\rho(r)$ in the order they are found. The restrictions on the parameters and/or coordinate result from the requirement that $A(r)$ and $B(r)$ must be both positive or both negative, depending on the case.

\subsection{Case ND (KS)} \label{ndsols}

We start with this case since it is the most physically relevant nonstandard case. We present five  solutions (two with the mathematical approach, two with the physical, one with mixed motivation). Because this case coincides with the Kantowski-Sachs spacetimes, and because of the simplicity of the starting ans\"{a}tze; we do not expect these solutions to be original; and we point out the first occurence of the solutions in the literature, when we can.\\

\noindent {\em Solution ND1.}

The simplest mathematical ansatz for (\ref{EEnd3}) is to take $B=$ Const. It can be any negative constant, but can be made equal to $-1$ by rescaling $t$. Then, 
\begin{eqnarray}
B & = & -1 {\rm   \;\;\;\;\;\;  (Ansatz)} \label{SolDn1B}\\
A & = & \frac{1}{1-C/r^{2}}     \label{SolDn1A} \\ 
\rho   & =  & \frac{C}{\kappa r^{4}}     \label{SolDn1rho}\\ 
p   & = & \frac{C}{\kappa r^{4}}     \label{SolDn1p}
\end{eqnarray}

$C$ must be positive and $r<\sqrt{C}$, so that this solution may serve as an interior solution. The equation of state turns out to be familiar, $p=\rho$, describing a stiff fluid\footnote{According to footnote 17 of~\cite{kantowskiTH}, this solution was first found in~\cite{thorneTH}, which we were unable to access.}.

Since $r$ is timelike (and if $u^{1}$ is taken to be negative), the increase of $p$ and $\rho$ with decreasing $r$ corresponds to compression as collapse progresses and their divergence at $r=0$ corresponds to the future singularity resulting from the collapse. Of course, positive $u^{1}$ will describe a spacetime region expanding from a past singularity.\\

\noindent {\em Solution ND2.}

The next simplest mathematical ansatz is to take $A$ to be a negative constant:
\begin{eqnarray}
A & = & {\rm -|constant|     \;\;\;\;\;\;  (Ansatz)} \label{SolDn2A}\\
B & = & - (C_{1} r^{\sqrt{1-A}}+C_{2} r^{-\sqrt{1-A}})^{2}     \label{SolDn2B} \\ 
p   & =  & -\frac{1}{\kappa r^{2}} \left(1-\frac{1}{A}\right)    \label{SolDn2p}\\ 
\rho   & = & \frac{1}{\kappa r^{2}} \frac{(A-1-2\sqrt{1-A}) C_{1}  r^{\sqrt{1-A}} + (A-1+2\sqrt{1-A}) C_{2}  r^{-\sqrt{1-A}}}
{A (C_{1}  r^{\sqrt{1-A}} + C_{2}  r^{-\sqrt{1-A}})}   \label{SolDn2rho}
\end{eqnarray}

Again,  $t$ can be rescaled to scale $B(r)$, therefore only the ratio $C_{2}/C_{1}$ is relevant, but we left $B(r)$ in this form to allow either constant to vanish. There is no restriction on $r$, and again we get compression towards infinite pressure at a future (or past) singularity.

The vanishing of either $C_{1}$ or $C_{2}$ leads to the proportional equation of state $p = w \rho$, but the range $-\frac{1}{3}<w<1$ is excluded. There is also a special value of $A$, for which the density does {\em not} diverge at the singularity, unless $C_{2}=0$. This occurs for $A-1+2\sqrt{1-A}=0$, that is, $A=-3$ ({\em Solution ND2*}), giving a density
\begin{equation}
\rho_{\rm ND2*}  = \frac{8}{3\kappa} \frac{C_{1} r^{2}}{C_{2} + C_{1} r^{4}}   \label{SolDn2*rho}
\end{equation}

The equation of state for $A=-3$ can be written as $\rho = - c_{3} \frac{p}{p^{2} +c_{4}} $  
with positive $c_{3}$ and $c_{4}$; not very familiar. For other values of $A$, the equation of state is even more complicated.\\

\noindent {\em Solution ND3.}

The simplest physical ansatz is the case of pressureless dust:
\begin{eqnarray}
p & = & 0 {\rm   \;\;\;\;\;\;  (Ansatz)} \label{SolDn3p}\\
A  & =  & \frac{1}{1-C/r}     \label{SolDn3A}\\ 
\rho & = & \frac{1}{\kappa r^{2} \left[ 1+\sqrt{\frac{C-r}{r}} \left(C_{1} - \tan^{-1}\sqrt{\frac{r}{C-r}}\right) \right] }     \label{SolDn3rho} \\ 
B   & = & -\left[ 1+\sqrt{\frac{C-r}{r}} \left(C_{1} - \tan^{-1}\sqrt{\frac{r}{C-r}}\right) \right]^{2}     \label{SolDn3B}
\end{eqnarray}

Again, $B$ has been scaled. $C$ must be positive and $r<C$, again giving an interior solution\footnote{This solution appears in the original Kantowski-Sachs paper~\cite{ks} as the $\epsilon=1$ case. (To see the equivalence, substitute for $t$ in terms of $\eta$ in their solution, and $r=C \cos^{2} \eta$ in ours).  The solution has no Schwarzschild limit, because its derivation excludes $\rho=0$. This subcase can easily be found from (\ref{EEnd1}) and (\ref{EEnd2}) (upon which (\ref{EEnd3}) needs to, and can be verified). \cite{ks} also gives the Schwarzschild limit separately, as the $\epsilon=0$ case.}, progressing towards a singularity at $r=0$. Interestingly, the form of $A$ is the same as in the (vacuum) Schwarzschild solution.     \\

\noindent {\em Solution ND4.}

One of the earliest solutions in case NS is that of an "incompressible" fluid, $\rho=$ const  \cite{schw-int}. Its mathematical analog  in case ND would be \mbox{$p=$ const.} This gives $F_{ND}=-\kappa p r^{3}/3 - C$. Then  (\ref{EEnd6}) becomes

\begin{equation}
\rho'    =   \frac{\kappa  r^{3} (\rho - p) -3C-4r }{2 r (r+\kappa p r^{3}/3 + C)} (\rho + p)
\end{equation}

We were unable to solve this equation for nonzero $C$, even with symbolic mathematics software. For $C=0$, however, an analytical solution can be found:
\begin{eqnarray}
p & = & \bar{p} = {\rm Const.   \;\;\;\;\;\;  (Ansatz)} \\
A  & =  & \frac{3}{\kappa \bar{p} r^{2}+3}     \\ 
\rho   & = &   \frac{2}{\kappa  r^{2} \left[\sqrt{-\frac{3+\kappa \bar{p} r^{2}}{3}} \left(c+\tan^{-1}\sqrt{-\frac{3}{3+\kappa \bar{p} r^{2}}}\right) - 1 \right] } - \bar{p}  \\ 
B   & = &  - \left[\sqrt{-\frac{3+\kappa \bar{p} r^{2}}{3}} \left(c+\tan^{-1}\sqrt{-\frac{3}{3+\kappa \bar{p} r^{2}}}\right) - 1 \right]^{2}
\end{eqnarray}

 Here $3+\kappa \bar{p} r^{2}$ must be negative, which means that $\bar{p}$ must be negative\footnote{We cannot take the $p \rightarrow 0$ limit here and recover the previous solution. Since we set $C=0$, $p \rightarrow 0$ means $A \rightarrow 1$, but $A$ should be negative.} and $r>\sqrt{-\frac{3}{\kappa \bar{p}}}$. Interestingly, the density diverges as $3+\kappa \bar{p} r^{2} \rightarrow 0$ while the pressure stays  constant.\\

\noindent {\em Solution ND5.}

The case of the "incompressible" fluid can be analyzed also for case ND, and is mathematically simpler:
\begin{eqnarray}
\rho & = & \bar{\rho} = {\rm Const.   \;\;\;\;\;\;  (Ansatz)} \label{SolDn4rho}\\
A  & =  & \frac{3}{\kappa \bar{\rho} r^{2}-1}     \label{SolDn4A}\\ 
p & = &  \bar{\rho} - \frac{4}{3 \kappa r^{2} }     \label{SolDn4p} \\ 
B   & = & - \frac{|C|}{r^{4} }    \label{SolDn4B}
\end{eqnarray}

One can also find this solution by looking for the expansionfree case, which gives $B=\frac{C}{r^{4}}$ \cite[Sect.15.6.1]{exsols}. Then eq. (\ref{EEnd3}) gives $A=-\frac{3}{1-C_{1}r^{2}}$, and (\ref{EEnd1}) and (\ref{EEnd2}) give $p$ and $\rho$.

For positive density there is a restriction on $r$, namely $\kappa \bar{\rho} r^{2}<1$, making this solution valid near the collapse singularity. What seems to be most interesting about this solution is that even an "incompressible" fluid can collapse (A given volume element shrinks along the $\theta$ and $\phi$ directions, but expands in the --now spacelike-- $t$ direction, conserving its volume). Pressure dominates over density near the singularity, one could even take $\bar{\rho}\rightarrow 0$, leaving only pressure. What better example to illustrate that pressure gravitates?

\subsection{Case TD} 

In this case also, $A(r)$ and $B(r)$ must be negative. All $A(r)$-$B(r)$-$p(r)$ triples given in the literature for case NS can be used where $A(r)$ and $B(r)$ are negative, but the source fluid is tachyonic and  $\rho(r)$ is different; in this sense they are ``new" solutions. Some simple examples are\\

\noindent {\em Solution TD1.}
\begin{eqnarray}
B & = & -1 {\rm   \;\;\;\;\;\;  (Ansatz)} \label{SolTD1B}\\
A & = & \frac{1}{1-Cr^{2}}     \label{SolTD1A} \\ 
p   & = & - C/\kappa     \label{SolTD1p}\\
\rho   & =  & - C/\kappa     \label{SolTD1rho}
\end{eqnarray}
The condition on $r$ is $C r^{2}>1$. $A(r)$ has the same form as the Einstein Static Universe solution\footnotemark[1], and this solution could be derived from it \cite{improper}.\\

\noindent {\em Solution TD2.}
\begin{eqnarray}
A & = & {\rm -|constant|     \;\;\;\;\;\;  (Ansatz)} \label{SolTD2A}\\
B & = &  -  (C_{1} r^{1+\sqrt{2-A}} + C_{2} r^{1-\sqrt{2-A}})^{2}  \label{SolTD2B} \\ 
p   & =  & \frac{1}{\kappa A r^{2}} \frac{(-A+3+2\sqrt{2-A}) C_{1} r^{\sqrt{2-A}}+(-A+3-2\sqrt{2-A}) C_{2} r^{-\sqrt{2-A}}}{C_{1} r^{\sqrt{2-A}} + C_{2} r^{-\sqrt{2-A}}  }    \label{SolTD2p}\\ 
\rho   & = & \frac{1}{\kappa A r^{2}} \frac{(A-5-4\sqrt{2-A}) C_{1} r^{\sqrt{2-A}}+(A-5+4\sqrt{2-A}) C_{2}  r^{-\sqrt{2-A}} }{C_{1} r^{\sqrt{2-A}} + C_{2} r^{-\sqrt{2-A}}  }   \label{SolTD2rho}
\end{eqnarray}

Like in solution ND2, $C_{1}$ and $C_{2}$ were both left in $B(r)$ to allow either constant to vanish; the vanishing of either one means $p=w\rho$ ($w$ between $\frac{1}{7} (-1 \pm 2 \sqrt{2})$ excluded); there is no restriction on $r$; $p$ diverges at $r=0$; there is a special value of $A$, -7, for which $\rho$ does {\em not} diverge. Then 
\begin{equation}
\rho_{TD2*}= \frac{24 r^{4}}{7 \kappa (C + r^{6})}.
\end{equation}\\

\noindent {\em Solution TD3.}

One ansatz that simplifies (\ref{EEtd6}) is $(\rho + 2p)=0$. Then,
\begin{eqnarray}
A  & =  & \frac{1}{1-C/r}     \label{Soltd3A}\\ 
p & = & \frac{8}{\kappa \left[ (2 r^{2}+5Cr-15 C^{2})+\sqrt{\frac{C-r}{r}} \left(C_{1}-15 C^{2} \tan^{-1}\sqrt{\frac{C-r}{r}}\right) \right] }        \label{SolTd3p} \\ 
\rho & = & -2p \\
B   & = & - r_{1}^{-4} \left[ (2 r^{2}+5Cr-15 C^{2})+\sqrt{\frac{C-r}{r}} \left(C_{1}-15 C^{2} \tan^{-1}\sqrt{\frac{C-r}{r}}\right) \right]^{2}     \label{SolTd3B}
\end{eqnarray}
with positive $C$ and $r<C$. \\

\noindent {\em Solution TD1'.}
\begin{eqnarray}
p   & = & \bar{p}= {\rm Const  \;\;\;\;\;\;  (Ansatz)}     \label{SolTd1p}\\
A & = & \frac{1}{1+\kappa \bar{p} r^{2}}     \label{SolTD0A} \\ 
B & = & {\rm Const } = -1  \label{SolTD0B}\\
\rho   & =  & \bar{p}     \label{SolTD11rho}
\end{eqnarray}
Since $A$ must be negative, $p$ must be negative, therefore this solution is identical to solution TD1. So, unlike in case ND(KS), constant $p$ leads necessarily to constant $\rho$.\\

\subsection{Case TS}

Static solutions supported by tachyonic fluids seem to be new. Here we must have positive  $A$ and $B$; as long as this can be satisfied, $A$, $B$, $p$ triples can be taken over from case ND(KS).\\

\noindent {\em Solution TS1.}

We first take $B=$ Const, then take over $A$ and $p$ from Solution ND1:
\begin{eqnarray}
B & = & 1 {\rm   \;\;\;\;\;\;  (Ansatz)} \label{SolTs1B}\\
A & = & \frac{1}{1-C/r^{2}}     \label{SolTs1A} \\ 
p   & = & \frac{C}{\kappa r^{4}}     \label{SolTs1p}\\
\rho   & =  & \frac{-3C}{\kappa r^{4}}     \label{SolTs1rho}
\end{eqnarray}

If $C$ is negative, there is no restriction on $r$; if $C$ is positive, we must have $r>\sqrt{C}$.\\

\noindent {\em Solution TS2.}

Next we take $A=$ Const, and take over $B$ and $p$ from Solution ND2:
\begin{eqnarray}
A & = & {\rm |constant|     \;\;\;\;\;\;  (Ansatz)} \label{SolTs2A}\\
B & = & (C_{1} r^{\sqrt{1-A}}+C_{2} r^{-\sqrt{1-A}})^{2}     \label{SolTs2B} \\ 
p   & =  & -\frac{1}{\kappa r^{2}} \left(1-\frac{1}{A}\right)    \label{SolTs2p}\\ 
\rho   & = & \frac{1}{\kappa r^{2}} \frac{(A-1+2\sqrt{1-A}) C_{1}  r^{\sqrt{1-A}} + (A-1-2\sqrt{1-A}) C_{2}  r^{-\sqrt{1-A}}}
{A (C_{1}  r^{\sqrt{1-A}} + C_{2}  r^{-\sqrt{1-A}})}   \label{SolTs2rho}
\end{eqnarray}

There is no restriction on $r$, both $\rho$ and $p$ diverge at the origin of $r$ for all values of $A$, except the special value 1, for which\footnote{At first sight, this solution is only valid for \mbox{$A<1$}.  For \mbox{$A>1$} one can go through the solution afresh and find \mbox{$B = B_{0} \cos^{2}[\sqrt{A-1} \ln(Cr)]$} and \mbox{$\rho = \frac{1}{\kappa r^{2}} \left(A-1-2\sqrt{A-1} \tan[\sqrt{A-1} \ln(Cr)] \right)$}. But these one can also find from (\ref{SolTs2A})-(\ref{SolTs2rho}) by expanding the exponentials of imaginary $\sqrt{1-A}$ in terms of trigonometric functions.} the solution degenerates into  {\em Solution TS2*}:
\begin{eqnarray}
A & = & 1 {\rm     \;\;\;\;\;\;  (Ansatz)} \label{SolTs2*A}\\
B & = & [\ln(r/r_{1})]^{2}     \label{SolTs2*B} \\ 
p   & =  & 0    \label{SolTs2*p}\\ 
\rho   & = & \frac{2}{\kappa r^{2} \ln(r/r_{1}) }   \label{SolTs2*rho}
\end{eqnarray}

\noindent {\em Solution TS3.}

Now we take $p=0$ case, giving same form of $A$ as in Solution ND3. But we cannot take over $B$ because $r-C$ must be positive. Then,
\begin{eqnarray}
p & = & 0 {\rm   \;\;\;\;\;\;  (Ansatz)} \label{SolTS3p}\\
A  & =  & \frac{1}{1-C/r}     \label{SolTS3A}\\ 
\rho & = & \frac{1}{\kappa r^{2} \left[ \sqrt{\frac{r-C}{r}} \ln\left(\frac{\sqrt{r}+\sqrt{r-C}}{\sqrt{r_{2}}}\right) -1\right] }     \label{SolTS3rho} \\
B   & = & \left[ \sqrt{\frac{r-C}{r}} \ln\left(\frac{\sqrt{r}+\sqrt{r-C}}{\sqrt{r_{2}}}\right) -1\right]^{2}     \label{SolTS3B}\\
\end{eqnarray}

If $C$ is negative, there is no restriction on $r$. If $C$ is positive, we must have $r>C$.\footnote{Because $p=0$, in the $C \rightarrow 0$ limit this solution should agree with Solution TS2*. Then, $A \rightarrow 1$, $\rho \rightarrow  \frac{1}{\kappa r^{2} \left[ \ln\left(2 \sqrt{r/r_{2}}\right) -1\right] }$, $B \rightarrow \left[ \ln\left(2\sqrt{\frac{r}{r_{2}}}\right) -1\right]^{2}$. The equivalence can be seen by multiplying this $B$ by 4 and defining $r_{2} = 4 r_{1}/e^{2}$.}\\

\noindent {\em Solution TS4.}

Now we try $p=\bar{p}=$ const. As in ND4, $F_{TS}=-\kappa \bar{p} r^{3}/3 - C$, but we can only solve the case $C=0$:
\begin{eqnarray}
p & = & \bar{p} = {\rm Const.   \;\;\;\;\;\;  (Ansatz)}  \label{SolTs4p} \\
A  & =  & \frac{3}{\kappa \bar{p} r^{2}+3}   \label{SolTs4A}   \\ 
\rho   & = & -  \frac{2}{\kappa  r^{2} \left[\sqrt{\frac{3+\kappa \bar{p} r^{2}}{3}} \left(c+\tanh^{-1}\sqrt{\frac{3}{3+\kappa \bar{p} r^{2}}}\right) - 1 \right] } - \bar{p}   \label{SolTs4rho} \\ 
B   & = &   \left[\sqrt{\frac{3+\kappa \bar{p} r^{2}}{3}} \left(c+\tanh^{-1}\sqrt{\frac{3}{3+\kappa \bar{p} r^{2}}}\right) - 1 \right]^{2}     \label{SolTs4B}
\end{eqnarray}
Here $3+\kappa \bar{p} r^{2}$ must be positive for $A$ to have correct sign, but  $\bar{p}$ must also be positive for $\tanh^{-1}\sqrt{\frac{3}{3+\kappa \bar{p} r^{2}}}$ to be real. Therefore there is no restriction on $r$.\\

\noindent {\em Solution TS5.}

We can take over $A$,  $B$,  $p$ from Solution ND5:
\begin{eqnarray}
A  & =  & \frac{3}{\kappa C_{1} r^{2}-1}     \label{SolTs5A}\\ 
B   & = & \frac{C}{r^{4} }    \label{SolTs5B}\\
p & = &  C_{1} - \frac{4}{3 \kappa r^{2} }     \label{SolTs5p} \\
\rho & = &  \frac{8}{3 \kappa r^{2} } - 3 C_{1}  \label{SolTs5rho}
\end{eqnarray}

The restriction on $r$ is  $\kappa C_{1} r^{2}>1$; so $C_{1}$ and $C$ must be positive. One can also arrive at this solution by taking $\rho + 2p =$ Const.\\

\subsection{Case NS} 

As stated in the beginning, this case is well-treated in the literature. One interesting solution is the analog of the $\bar{\rho} \rightarrow 0$ limit of Solution ND5:\\

\noindent {\em Solution NS1} (Kuch68 I):
\begin{eqnarray}
\rho & = & 0 {\rm   \;\;\;\;\;\;  (Ansatz)} \label{SolNSrho}\\
A  & =  & \frac{1}{1-C/r}     \label{SolNSA}\\ 
p & = & \frac{8}{\kappa \left[ (2 r^{2}+5Cr-15 C^{2})+\sqrt{1-C/r} \left(C_{1}+15 C^{2} \ln(\frac{\sqrt{r-C}+\sqrt{r}}{\sqrt{|C|}})\right) \right] }        \label{SolNSp} \\ 
B   & = & r_{1}^{-4} \left[ (2 r^{2}+5Cr-15 C^{2})+\sqrt{1-C/r} \left(C_{1}+15 C^{2}  \ln(\frac{\sqrt{r-C}+\sqrt{r}}{\sqrt{|C|}})\right)\right]^{2}     \label{SolNSB}
\end{eqnarray}
where if $C$ is positive, we must have $r>C$~\footnote{This solution was found in \cite{Kuch68I} and named Kuch68 I in~\cite{delgaty&lake} (To see the equivalence, square the argument of the $\ln$ above, and redefine the constants). Since its $A(r)$ is the same as that of the Schwarzschild exterior solution, it is obtainable from that by the transformation T2 of \cite{visseretal}. We can put any length into the root in the denominator of the argument of $\ln$ by redefining $C_{1}$. The choice was made for agreement with \cite{delgaty&lake} or \cite{visseretal} while allowing $C$ to be negative. Neither of \cite{delgaty&lake} or \cite{visseretal} or~\cite{Kuch68I} mention the $r>C$ restriction, although~\cite{Kuch68I} mentions a similar restriction for the solution named Kuch68 II in~\cite{delgaty&lake}.}. The $C=0$ case  is simple, and is valid for all $r$:\\

{\em Solution NS1*} (K-O III):
\begin{eqnarray}
\rho & = & 0 \label{SolNS*rho}\\
A  & =  & 1     \label{SolNS*A}\\ 
p & = & \frac{4}{\kappa r^{2}+C_{2}}          \label{SolNS*p} \\ 
B   & = & C_{3} ( \kappa r^{2}+C_{2})^{2}     \label{SolNS*B}
\end{eqnarray}
This solution\footnote{According to \cite{delgaty&lake}, this solution was found in \cite{K-O3}; but actually~\cite{Kuch68I} gives this as a special case of Kuch68 I. Since its $A(r)$ is the same as that of Minkowski metric, it is obtainable from that by the transformation T2 of \cite{visseretal}. Neither of \cite{delgaty&lake} or \cite{visseretal} or~\cite{Kuch68I} mention that this solution is special because it is free of the coordinate range restriction.} demonstrates, like solution ND5, that pressure gravitates: There is no density to gravitate, and pressure must obviously be providing the gravitational attraction against its own repulsion to keep the system static (In the $\rho \rightarrow 0$ limit of Solution ND5, the pressure was negative). For positive central pressure $p_{0}$, the solution is regular, for negative $p_{0}$, there is a singularity at $r=\frac{2}{\sqrt{- \kappa p_{0}}}$.

\section{Conclusions} 

In this work, we asked how far one can go with the ansatz (\ref{ansatz}), together with the assumption that the source is a perfect fluid. That ansatz is usually used when looking for static spherically symmetric solutions.

We pointed out that the requirement of correct signature means that the two metric functions $A(r)$ and $B(r)$ in (\ref{ansatz}) must be both positive or both negative. The natural followup question is how these two possibilities should be interpreted. The answer to this question led to four cases, depending on if the source four-velocity is directed along the $t$-coordinate or along the $r$-coordinate, and if the source four-velocity is timelike or spacelike. So we found that apart from the standard case corresponding to static perfect fluid solutions, the ansatz (\ref{ansatz}) can cover the Kantowski-Sachs class of solutions (after relabeling the radial and time coordinates) and two cases with tachyonic fluid as source. For each case, we wrote down the Einstein Equations, and derived Oppenheimer-Volkoff-like formalisms, that is, equations suited for finding solutions starting from an equation of state. We derived some simple solutions as examples.

Some papers pointed out before that in some dynamic spacetimes the source fluid may be tachyonic, but here we found that this feature is generic in two of the four cases covered by ansatz (\ref{ansatz}). On the other hand, it seems not to have been noticed before that once tachyonic source is considered, static spacetimes are also possible.

The four classes of solutions found here are physically quite distinct and should be considered local solutions, especially when there is a coordinate range restriction. We did not discuss in this work any possible patching together of solutions of different classes. It may seem that sometimes a solution has regions of different classes by virtue of metric functions switching sign at some $r$-value, which would form an apparent horizon. But then, the nature (normal vs. tachyonic) or the four-velocity (directed along $t$ vs. directed along $r$) of the source fluid would have to be wildly different on the two sides of the horizon (unless there is no source fluid, as in the Schwarzschild case). 

On the other hand, a spherically symetric black hole is just such a spacetime, one which features a horizon at some $r=R$, where for $r>R$, $r$ is spacelike and for $r<R$, $r$ is timelike. In other words,  it is a composite of a class NS or TS spacetime (for $r>R$), and a class ND or TD spacetime (for $r<R$). Because these classes are physically quite distinct, {\em spherically symmetric black hole solutions with perfect fluid source seem to be impossible if one requires the metric elements and source fluid properties to be given by one function of $r$ each, across the whole range of $r$.} This is valid for all equations of state, dark energy ($\rho+3p<0$), phantom energy ($\rho+p<0$) and negative energy density ($\rho<0$) included; but is not valid for non-perfect-fluid cases (e.g. Reissner-Nordstr\"{o}m), and of course, vacuum (Schwarzschild). So, a spherically symmetric perfect-fluid black hole would either have  its spacetime patched together from solutions of different classes, or have a discontinuity of some property of the source fluid on the horizon, or be given by an ansatz more complicated than (\ref{ansatz}).\\

\noindent {\bf Acknowledgements}\\

We would like to thank to \"{O}. G\"{u}rtu\u{g}, T. Rador and M. \"{O}zbek for stimulating and helpful discussions, and A.Kargol for help with access to some references. This work was partially supported by Grant No. 06B303 of the Bo\u{g}azi\c{c}i University Research Fund.

\end{document}